\documentclass[a4paper]{article}
\usepackage{graphicx}
\usepackage{onecolceurws}

\usepackage{hyperref}
\usepackage{amssymb}
\usepackage{mathtools}

\usepackage{color}
\definecolor{keywordcolor}{rgb}{0.7, 0.1, 0.1}   
\definecolor{commentcolor}{rgb}{0.4, 0.4, 0.4}   
\definecolor{symbolcolor}{rgb}{0.0, 0.1, 0.6}    
\definecolor{sortcolor}{rgb}{0.1, 0.5, 0.1}      

\usepackage{listings}

\providecommand{\keywords}[1]{\textbf{\textit{Keywords---}} #1}

\providecommand{\MSC}[1]{\textbf{2010 MSC:} #1}

\title{Elements of Differential Geometry in Lean \\
A Report for Mathematicians}

\author{
Anthony Bordg \\ Dept. of Computer Science and Technology\\
                University of Cambridge, UK \\  apdb3@cam.ac.uk
\and
 Nicol\`{o} Cavalleri \\ Dept. of Computer Science and Technology\\
 University of Cambridge, UK \\ nc555@cam.ac.uk
}

\institution{}

\begin{document}
\maketitle

\begin{abstract}
We report on our experience formalizing differential geometry with mathlib, the Lean mathematical library. Our account is geared towards geometers with no knowledge of type theory, but eager to learn more about the formalization of mathematics and maybe curious enough to give Lean a try in the future. To this effect, we stress the possibly surprising difference between the formalization and its pen-and-paper counterpart arising from Lean's treatment of equality. Our three case studies are Lie groups, vector bundles and the Lie algebra of a Lie group. 
\end{abstract}
\vskip 32pt

\MSC{53-04, 03B35, 68V20}


\subsubsection*{Supplement Material}

Our formal development is now part of mathlib\footnote{See \url{https://github.com/leanprover-community/mathlib/blob/master/src/geometry/manifold/algebra/lie_group.lean} for Lie groups, \url{https://github.com/leanprover-community/mathlib/blob/master/src/topology/vector_bundle.lean} for topological vector bundles and \url{https://github.com/leanprover-community/mathlib/blob/master/src/geometry/manifold/algebra/left_invariant_derivation.lean} for the Lie algebra of a Lie group.}.

\section{Introduction}

The logical system underlying the Lean theorem prover is known as an intensional dependent type theory \cite{Streicher}. We will focus on one of the most subtle notions in such a system: equality. To learn more on Lean's dependent type theory, the reader should refer to \cite{Theorem-Proving-Lean}. 

\subsection{Equality in Lean}

The singular ``equality'' in the title actually hides a plural. Indeed, there are multiple notions of equality in Lean, most notably the so-called \textit{definitional equality} and the so-called \textit{propositional equality}. Some traces of this distinction can be found in traditional mathematics with the  ``defined to be equal to'' symbol $\coloneqq$ and with the non-trivial equalities stated as propositions which have to be proved. Unsurprisingly, two terms that are definitionally equal in Lean can be used interchangeably. In order to be able to refer to the equality between two terms within  Lean's syntax, there exists a second kind of equality, namely the propositional equality. However, if in pen-and-paper mathematics two expressions which have been proved equal can now be used interchangeably, blurring the above distinction, in Lean these two notions of equality are really distinct. Indeed, in Lean two terms that are propositionally equal may not be definitionally equal. For instance, the equality $n + 0 = n$ is true by definition in Lean, while $0 + n = n$ is not and as a proposition requires a proof. Of course, definitional equality depends on implementation, something that mathematicians may not be used to take into account. 
After some work, implementation details can to some extent be abstracted away from the user, but they clearly affect the experience of the developer building an interface for a given mathematical structure. In pen-and-paper mathematics, mathematical structures are ethereal specifications without implementations and in the example above natural numbers form a totally ordered commutative semiring in which the definition of addition as such does not matter.
We will see what kind of unexpected challenges these distinctions pose for the translation of pen-and-paper mathematics into Lean.

\subsection{Differential Geometry in Mathlib}

Manifolds in Lean are smooth manifolds locally modeled on a model with corners over a nondiscrete normed field $\Bbbk$, \textit{i.e.} a field equipped with a valuation, namely a real-valued map $\varphi$ satisfying $\varphi(x) = 0$ if and only if $x = 0$ and
\[
\varphi(xy) = \varphi(x) \varphi(y), \quad \varphi(x + y) \leq \varphi(x) + \varphi(y)
\] 
for all $x,y \in \Bbbk$, and in which there exists an element $x$ with $\varphi(x) < 1$.
The field $\Bbbk$ can be $\mathbb{R}$ together with the absolute value for instance and in this case in pen-and-paper mathematics $n$-dimensional manifolds with corners are locally modeled on $[0, \infty)^l \times \mathbb{R}^{n-l}$ for some $l$ with $0 \leq l \leq n$. If smooth manifolds without boundary and manifolds with boundary are well known,  manifolds with corners are comparatively less known. Obviously, manifolds with corners generalize both boundaryless manifolds (case $l \coloneqq 0$) and manifolds with boundary (case $l \coloneqq 1$). While a nondiscrete normed field does not add much complication and allows to treat both real and complex manifolds at the same time, the set-up of manifolds with corners in Lean requires some formal machinery as we will briefly see below and using them may be cumbersome for the casual user.  
Beyond some basic material, as of June 2021 the Lean library for differential geometry lacks more advanced material, like de Rham cohomology and Riemannian geometry, that could stress-test the design choices made. It will be interesting to follow the evolution of this library.

\section{Case Studies}

Smooth manifolds in Lean as formalized by S\'ebastien Gou\"ezel are modeled on a model with corners, denoted \lstinline{I : model_with_corners $\Bbbk$ E H} in Lean. Given a model vector space $E$ and a model space $H$, a model with corners is defined as a map $I: H \rightarrow E$ embedding nicely $H$ into $E$ with respect to which the transition functions, which are maps from $H$ to $H$, are smooth. In the case where $\Bbbk \coloneqq \mathbb{R}$, one example of a model with corners is the embedding of $[0, \infty)^n$ into $\mathbb{R}^n$. In the trivial case where the map $I$ is the identity denoted \lstinline{id}, embedding $E$ into itself, there is in Lean a trivial model with corners denoted \lstinline{model_with_corners_self $\Bbbk$ E}. 
In the case of boundaryless manifolds we shall see the complications arising from Lean's treatment of equality. In this case one would expect to work only with the trivial model with corners, but this is not the case. Indeed, assume that we have two model vector spaces $E$ and $F$. There are two natural models with corners on $E \times F$, namely the identity \lstinline{id} and the product \lstinline{id $\times$ id}, namely the map  \lstinline{(λp : E × F, (p.1, p.2))}, where \lstinline{p.1} and \lstinline{p.2} denote the first and the second projection respectively. These two maps are not definitionally equal, hence we have to indicate to Lean which one we want to use. More precisely, the map \lstinline{id} arises from considering $E \times F$ itself as a boundaryless manifold, while the map \lstinline{id $\times$ id} arises from considering $E \times F$ as the product of two (boundaryless) manifolds, this product being denoted \lstinline{model_prod} in the library. This exemplifies well the kind of subtleties involved with the equality in Lean.  As a result, for boundaryless manifolds a dedicated property\footnote{Technically, something called a \textit{type class} in Lean's type theory.} had to be introduced by Gou\"ezel. This property, simply ensuring that a manifold has no boundary, has an unspecified model with corners to maximize the applicability of its theorems.   

\subsection{Lie Groups}

In this subsection we will see in the case of Lie groups the consequences of the design choices for smooth manifolds outlined above. 
First, given the set-up of manifolds in Lean, a Lie group in Lean is a priori a manifold with corners.
Second, in the case of Lie groups one would expect to use the trivial model with corners given by the identity map \lstinline{id}, namely the model \lstinline{model_with_corners_self}. However, the product of two manifolds $M$ and $M'$ in the library requires to use \lstinline{model_prod H H'} as the model space of $M \times M'$, where $H$ and $H'$ are the respective model spaces of $M$ and $M'$. As a consequence,  using the trivial model with corners for formalizing Lie groups would not make Lie groups stable under products, \textit{i.e.} it would not be possible to prove that the product of two Lie groups is again a Lie group.
Indeed, assume that $M \coloneqq G$ and $M' \coloneqq G'$ are two Lie groups, each endowed with the trivial model with corners, hence $H$ (resp. $H'$) is $E$ (resp. $E'$) and both $I$ and $I'$ are \lstinline{id}, then the product $G \times G'$ would be endowed with the model with corners \lstinline{id $\times$ id}, which is not definitionally equal to \lstinline{id} as explained previously,  so $G \times G'$ would not be a Lie group if Lie groups were defined using the trivial model with corners.
Thus, the definition of a Lie group has to be generalized, maybe awkwardly, allowing a model with corners where the model space and the model vector space can be distinct. 
We give below the code for our definition of Lie groups for the curious reader who wants to know what Lie groups look like in Lean, but we do not expect them to be able to make sense of the code.  

\begin{lstlisting}
@[ancestor has_smooth_mul, to_additive]
class lie_group {$\Bbbk$ : Type*} [nondiscrete_normed_field $\Bbbk$]
  {H : Type*} [topological_space H]
  {E : Type*} [normed_group E] [normed_space $\Bbbk$ E] (I : model_with_corners $\Bbbk$ E H)
  (G : Type*) [group G] [topological_space G] [charted_space H G]
  extends has_smooth_mul I G : Prop :=
(smooth_inv : smooth I I (λ a:G, a⁻¹))
\end{lstlisting}
We then proceed to prove without any difficulty that the product of two Lie groups is again a Lie group.

\subsection{Vector Bundles}

One could define a vector bundle in Lean as a surjective continuous map $\pi: E \rightarrow B$ satisfying standard conditions \cite[chap. 3]{Spivak}. However, in case of a naive implementation of this definition the fibers of the tangent bundle would not be definitionally equal to the tangent spaces, only isomorphic to them. This is perfectly acceptable for the working mathematician, but now the Lean developer would have two isomorphic copies of the tangent space at a given point. As a consequence, our developer would have to transport theorems about these copies along an isomorphism, something which is never carried out explicitly on paper, since mathematicians treat isomorphic objects as equal. Thus, following a suggestion of Mario Carneiro we defined instead a topological vector bundle as a map \lstinline{E : B → Type*}, \textit{i.e.} a family of types, the fibers, indexed by a base space $B$, such that each $E(x)$ is a topological vector space and around every point there is a local trivialization which is linear in the fibers. This definition of vector bundles has the merit of providing a more direct way of talking about fibers, avoiding to deal with fibers only isomorphic to tangent spaces. Thus, topological vector bundles translate into Lean as follows.

\begin{lstlisting}
class topological_vector_bundle (R : Type*) {B : Type*} (F : Type*) (E : B → Type*)
[semiring R] [∀ x, add_comm_monoid (E x)] [∀ x, module R (E x)]
[topological_space F] [add_comm_monoid F] [module R F]
[topological_space (total_space E)] [topological_space B] : Prop :=
(inducing [] : ∀ (b : B), inducing (λ x : (E b), (id ⟨b, x⟩ : total_space E)))
(locally_trivial [] : ∀ b : B, ∃ e : topological_vector_bundle.trivialization R F E, b ∈ e.base_set)
\end{lstlisting}

\subsection{The Lie Algebra of a Lie Group}

\subsubsection{Derivations of the Algebra of Smooth Functions on a Manifold}

We already touched on the isomorphism-as-identity problem in the previous section. We will give one additional incarnation of this problem to exemplify how pervasive it is in the formalization of mathematics with proof assistants. Given a real smooth manifold $M$, it is known\footnote{\textit{Cf.} \url{https://ncatlab.org/nlab/show/derivations+of+smooth+functions+are+vector+fields}} that $\text{Der}(C^{\infty}(M))$, the real vector space of derivations of the algebra $C^{\infty}(M)$ of smooth functions on $M$, is isomorphic to $\Gamma^{\infty}(M, T M)$, the real vector space of smooth vector fields on $M$.
\[
\Gamma^{\infty}(M, T M) \cong \text{Der}(C^{\infty}(M))
\]
This isomorphism is an isomorphism in the category of Lie algebras over $\mathbb{R}$.
Since derivations of the algebra of smooth functions is a more algebraic theory, it is particularly well suited for the dependent type theory of Lean and we chose to follow this approach as shown in the snippet of code below.
\begin{lstlisting}
@[protect_proj]
structure derivation (R : Type*) (A : Type*) [comm_semiring R] [comm_semiring A]
  [algebra R A] (M : Type*) [add_cancel_comm_monoid M] [module A M] [module R M]
  [is_scalar_tower R A M]
  extends A → [R] M :=
(leibniz' (a b : A) : to_fun (a * b) = a • to_fun b + b • to_fun a)
\end{lstlisting}
However, in some situations one might prefer to work with vector fields and even once the above isomorphism has been proved, it would remain to transfer along the isomorphism properties from one side to the other and this would not be handled automatically by Lean. This is a significant difference between Lean and the practice of pen-and-paper mathematics where isomorphic objects are treated as equal. It is worth noting that in some dependent type theories like Voevodsky's Univalent Foundations \cite{HoTTBook} this problem has been eased\footnote{Lean is not compatible with the Univalent Foundations since its third version, Lean 3.}. Indeed, in the Univalent Foundations, given two objects of some category of structured sets there is an equivalence between the type of their equalities and the type of their isomorphisms \cite{coquand2013isomorphism}, ensuring that properties can always be transferred between isomorphic objects by invoking appropriate theorems. Despite this advantage, the treatment of equality in the Univalent Foundations is even more subtle than in Lean, since the equality between two objects is not a mere proposition but it has the structure of an $\infty$-groupoid. 

\subsubsection{Left-invariant Derivations and the Lie Algebra of a Lie Group}

We formalized in Lean the property of being left-invariant for derivations which is analogous to the property of being left-invariant for vector fields. In other words, this property of being left-invariant commutes with the aforementioned isomorphism, \textit{i.e.} a derivation is left-invariant if and only if its image under this isomorphism is a left-invariant vector field. We encountered a notable subtlety during the formalization of left-invariant derivations that we shall explain. Let $G$ be a Lie group and $g$ be any element of $G$. The symbol $L_g$ will denote the left translation  of $G$ and $e$ its identity. Actually, $L_g e$ and $g$ are not definitionally equal in Lean, only propositionally equal. As a result, the tangent spaces $T_{L_g e} G$ and $T_g G$ are not definitionally equal, only propositionally equal. This mismatch is not only a problem for defining the property of being left-invariant, but it could also be a problem in the future if one wants to prove for instance that every Lie group is parallelizable. Indeed, to prove that every Lie group is parallelizable, one will have to move around a basis of $T_e G$ using the left translations $L_g$'s to get a basis on each vector space $T_g G$, hence one will also have to address the above mismatch in this case. At this point we have to tell the reader a last surprising fact. There are not only two kinds of equalities in Lean as presented  in the introduction, but there is a third equality, the so-called \textit{heterogeneous equality} (heq), which can be used to handle the kind of mismatches described above, \textit{e.g.} between $T_{L_g e} G$ and $T_g G$. However, heq, denoted \lstinline{==} in Lean syntax, despite being an equivalence relation is strictly weaker than propositional equality and has notorious limitations. For instance, given two functions $f$ and $g$ and an element $x$ in their domain, \lstinline{f == g} does not necessarily imply \lstinline{f x == g x} \cite[section 3]{CongruenceClosureITT}. As a consequence, the use of heq often gives rise to some difficulties informally referred to as ``heq hell'' in the Lean community. Thus, we decided not to use heq and Oliver Nash suggested another way around our problem. Briefly, we introduced a new differential for derivations, an heterogeneous differential, which takes as arguments not only a smooth function $f$ and a point $x$ on the given manifold, but also a point $y$ and a proof of equality between $f(x)$ and $y$. This enables us to replace $T_{L_g e} G$ with $T_g G$ everywhere and our problem vanishes. This heterogeneous differential has been used only for derivations so far and a more standard, homogeneous, differential is used throughout the library. Last, we eventually proved that the left-invariant derivations form a Lie algebra. In particular, in the real case the left-invariant derivations are an implementation of the Lie algebra of a Lie group. 

\begin{lstlisting}
instance : lie_algebra $\Bbbk$ (left_invariant_derivation I G) :=
{ lie_smul := λ r Y Z, by { ext1, simp only [commutator_apply, map_smul, smul_sub, coe_smul,
              pi.smul_apply] } }
\end{lstlisting}      

\section{Concluding Thoughts}

There are hints that the approach taken for formalizing manifolds in Lean is scalable, though much remains to be done to extend the library which currently lacks an array of nontrivial examples for the notions already implemented. Regarding Lean itself we have to point out that the level of automation is not on a par with what most mathematicians would expect. Moreover, the fact that every term belongs to a type creates some difficulties that a mathematician will hardly suspect. Indeed, if Lean type class inference mechanism allows to see automatically a Lie group as a group or a Lie group as a smooth manifold, one needs the so-called coercions to see for instance a smooth map as a map. These coercions clutter the code and render it unnecessarily verbose from the mathematician's perspective always equating an object with its image under a forgetful functor or a canonical inclusion. A mathematician may also regret the limited support for notations that makes the code verbose and hard to decipher as well as the absence of \LaTeX\  support to comment extensively the code with formulas or diagrams. Also, given its sophisticated type theory Lean has problably a steep learning curve for most mathematicians and the cryptic messages given in the user interface whenever an error occurs are not of much help to the beginner. Finally, if mathematicians think they know what equality is, implementing their favorite mathematical structures in Lean may force them to revise this position.

\subsubsection*{Acknowledgements}

This work was supported by the ERC Advanced Grant ALEXANDRIA (Project GA 742178) and the Cambridge Mathematics Placements Programme from the Faculty of Mathematics of the University of Cambridge. We thank the maintainers of mathlib, as well as Oliver Nash, who took an active part in reviewing our code after it was submitted to mathlib. Among the mathlib maintainers we thank in particular S\'ebastien Gou\"ezel. We also thank one anonymous reviewer for their relevant remarks.


\begin{thebibliography}{Avi21}

\bibitem[Str93]{Streicher}
Thomas Streicher. 
\newblock Investigations into intensional type theory. 
\newblock Habilitiation Thesis, Ludwig Maximilian Universität, 1993.

\bibitem[Avi21]{Theorem-Proving-Lean}
J. Avigad, L. de Moura and  S. Kong.
\newblock Theorem Proving in Lean.
\newblock {\em Release 3.23.0}, Jan 2021.
\newblock \url{https://leanprover.github.io/theorem_proving_in_lean/}.

\bibitem[Spi99]{Spivak}
Michael Spivak.
\newblock A Comprehensive Introduction to Differential Geometry, Volume 1.
\newblock Publish or Perish Inc., 3rd edition, 1999.

\bibitem[Uni13]{HoTTBook}
The Univalent Foundations Program.
\newblock Homotopy Type Theory: Univalent Foundations of Mathematics.
\newblock Institute for Advanced Study, 2013.
\newblock \url{https://homotopytypetheory.org/book}

\bibitem[Coq13]{coquand2013isomorphism}
T. Coquand and N. A. Danielsson.
\newblock Isomorphism is equality.
\newblock \textit{Indagationes Mathematicae}, 24(4), pp.1105-1120, 2013.

\bibitem[Sel16]{CongruenceClosureITT}
Daniel Selsam and Leonardo de Moura. 
\newblock Congruence Closure in Intensional Type Theory. 
\newblock Proceedings of the 8th International Joint Conference on Automated Reasoning, Volume 9706, pp. 99-115, Springer-Verlag, Berlin, Heidelberg, 2016.

\end{thebibliography}
\end{document}